\documentclass[aps,prl,twocolumn,amsmath,amssymb,superscriptaddress,longbibliography]{revtex4-1}
\usepackage[english]{babel}
\usepackage{amsmath,amssymb,bm,graphicx,braket}
\usepackage[colorlinks=true,citecolor=blue,linkcolor=blue,urlcolor=blue]{hyperref}
\graphicspath{{pict/}{}}

\begin{document}


\title{Excitation-Selective Control of Protected Subradiant Antiresonances}

\author{Jing Tang}
\affiliation{School of Physics and Optoelectronic Engineering,
Guangdong University of Technology, Guangzhou 510006, China}
\affiliation{Guangdong Provincial Key Laboratory of Sensing Physics
and System Integration Applications, Guangdong University of Technology,
Guangzhou 510006, China}

\author{Yuangang Deng}
\email{dengyg3@mail.sysu.edu.cn}
\affiliation{Guangdong Provincial Key Laboratory of Quantum Metrology
and Sensing $\&$ School of Physics and Astronomy,
Sun Yat-Sen University, Zhuhai 519082, China}

\date{\today}

\begin{abstract}
Subradiant cavity antiresonances exhibit narrow spectral features, yet controlling their nonlinear response without perturbing the protected resonance remains challenging. Here we demonstrate excitation-selective control of a subradiant antiresonance in a driven two-emitter cavity-QED system. An auxiliary collective channel leaves the antisymmetric single-excitation state unaffected while shifting the doubly excited state, introducing tunable excitation anharmonicity without altering the linear antiresonance. This separation enables control of two-photon excitation pathways and photon correlations. Independently, a spatial coupling phase tunes optical access to the subradiant state without changing its intrinsic decay rate. We find that the interferometric phase information per incident photon is optimized when cavity-induced readout approximately matches intrinsic subradiant decay, while steady-state atomic entanglement peaks at an intermediate coupling phase. We further extend phase-selective optical control to collective subradiant modes in correlated emitter arrays. These results establish collective symmetry and spatial mode matching as distinct controls of nonlinear photon correlations, optical phase information, and atomic entanglement.
\end{abstract}

\maketitle

{\em Introduction.---}Interference between collective radiative pathways gives rise to superradiance and subradiance with decay rates markedly different from those of isolated emitters \cite{Dicke1954,Lehmberg1970,Gross1982,Ficek2002}. Subradiant states support long-lived optical coherence, but their weak radiative coupling often limits optical excitation and readout \cite{Ostermann2013,Ostermann2014,Filipp2011,Guerin2016,Facchinetti2016,Kornovan2019,Ferioli2021,ZhangMolmer2019,Eltohfa2025}. Ordered emitter arrays offer spatial control over collective optical modes, enabling cooperative scattering, selective radiance, and long-lived collective excitations \cite{Jenkins2012,Jenkins2013,Bettles2015,Bettles2016,Shahmoon2017,Perczel2017,AsenjoGarcia2017,Rui2020,Ballantine2020a,Ballantine2021,Srakaew2023,RubiesBigorda2022}. Their collective response is sensitive to geometry, atomic motion, and disorder \cite{Masson2020,RubiesBigorda2022PRA,Gjonbalaj2024,Yadav2024,Olmos2025}. Cavity and nanophotonic platforms provide additional control over collective excitations and their radiative properties \cite{Ritsch2013,Reimann2015,Casabone2015,Douglas2015,Pichler2015,Solomons2023,Yan2023,Hotter2023,Zhou2025,Kim2025,MassonAsenjo2020}. In particular, cavity antiresonances provide a sensitive probe of the emitter response \cite{Sames2014}, while coupling to subradiant collective modes can yield narrow antiresonances with enhanced cooperativity and steep phase dispersion \cite{Plankensteiner2017}. These developments raise a question beyond linear spectroscopy: can higher-excitation dynamics be controlled without perturbing a subradiant antiresonance?

This question is nontrivial because subradiant protection, optical accessibility, and nonlinear response involve distinct aspects of collective dynamics. A long-lived excitation may couple weakly to an optical probe, whereas stronger cavity coupling can deepen the antiresonance and suppress transmission. Thus, steep phase dispersion alone does not guarantee greater phase information per incident photon, an important consideration for collective optical metrology \cite{Meiser2009,Bohnet2012,Ostermann2013,Barberena2020,Norcia2016,Norcia2018,Liu2020}. Beyond linear response, photon correlations depend on higher-excitation manifolds and their interference pathways, which are not determined by the single-excitation spectrum \cite{Birnbaum2005,Flayac2017,Snijders2018,ZhangMolmer2019}. Strong photon bunching can arise when single-photon excitation is suppressed relative to higher-order processes, as recently explored in cavity-mediated subradiance \cite{Kim2025}. Collective dissipation, spatially selective coupling, and cavity-mediated interactions provide routes to controlling dark states and collective excitation dynamics \cite{Gegg2018,Shankar2021,Pineiro2022,Qiao2022,Zanner2022,Patti2021,RubiesBigorda2022,Sundar2024,Sundar2024PRA}. Related reservoir-engineering approaches have also enabled the preparation and stabilization of entangled states \cite{Krauter2011,Shankar2013,Vivas2024PRL,Vivas2024PRR,Ron2024}. The central challenge is to modify higher-excitation pathways without disturbing the collective state responsible for the linear antiresonance.

Here we demonstrate excitation-selective control in a driven two-emitter cavity-QED system. The emitters couple to a primary cavity with a tunable relative phase, while an auxiliary cavity mediates spin-exchange interactions (SEI) and collective dissipation. The antisymmetric single-excitation state is dark to the auxiliary channel, whereas the symmetric single-excitation and doubly excited states remain coupled. The SEI shifts the latter states while leaving the antisymmetric state unaffected, introducing a tunable anharmonicity into the excitation ladder. When the primary cavity selectively addresses the antisymmetric state, the linear antiresonance remains unchanged, while the modified two-excitation spectrum allows photon correlations to be tuned. Independently, the primary-cavity coupling phase controls optical access to the subradiant state without altering its intrinsic decay rate. We further investigate the resulting phase-sensitive optical response and steady-state atomic entanglement, and extend phase-selective optical control to collective subradiant modes in correlated emitter arrays.

\begin{figure}[t]
\centering
\includegraphics[width=\columnwidth]{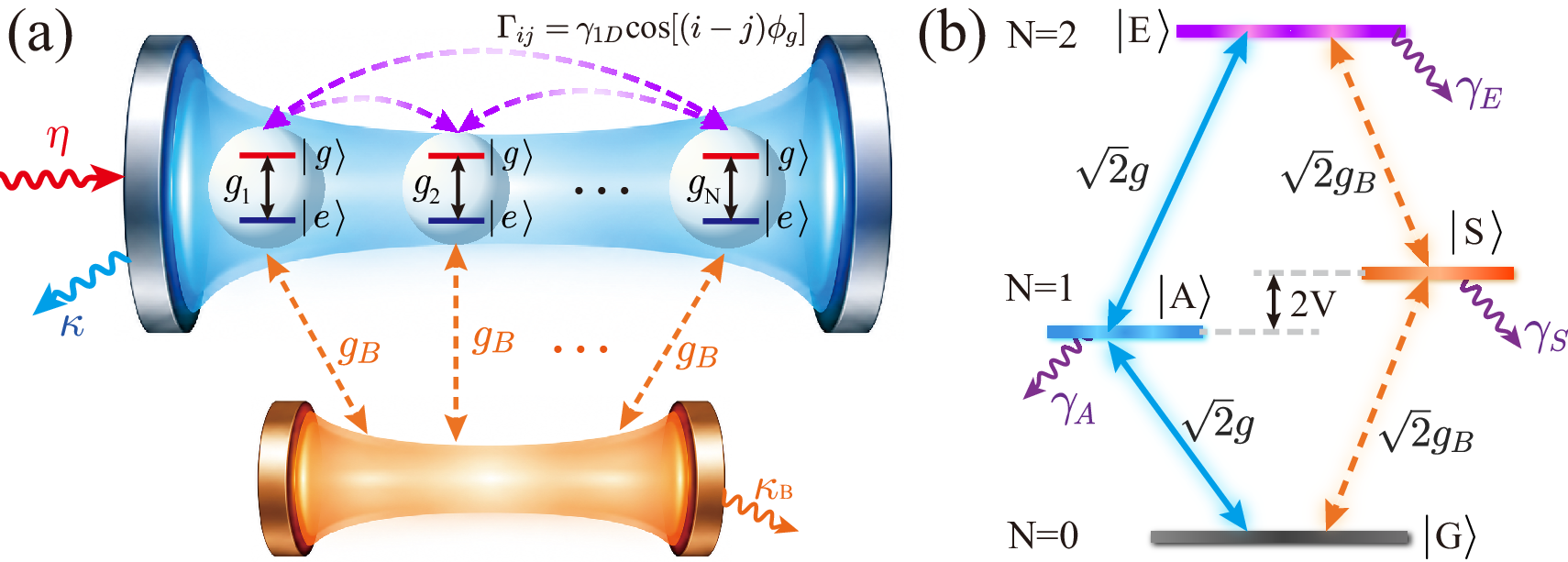}
\caption{\textbf{Cavity-QED model and excitation-selective protection.} (a) An array of $N$ two-level emitters coupled to a driven primary cavity and a far-detuned auxiliary cavity, with spatially correlated atomic decay. (b) Collective-state level structure for $N=2$ at $\phi=\pi$. The primary cavity (blue arrows) couples the $|G\rangle\leftrightarrow|A\rangle\leftrightarrow|E\rangle$ ladder, while the auxiliary channel (orange dashed arrows) couples the symmetric sector, leaving $|A\rangle$ protected but modifying the two-excitation response.}
\label{fig:model}
\end{figure}

{\em Model and excitation-selective protection.---}We consider $N$ identical two-level emitters coupled to a driven primary cavity and an auxiliary cavity [Fig.~\ref{fig:model}(a)]. We first establish the excitation-selective mechanism for two emitters and then extend the linear optical response to correlated emitter arrays. The primary cavity mode $\hat a$ couples to the emitters with position-dependent strengths $g_j=g\cos[(j-1)\phi]$, where $\phi$ is the relative coupling phase, while the auxiliary mode $\hat b$ couples uniformly with strength $g_B$. In the dispersive regime, adiabatic elimination of the auxiliary cavity induces an effective infinite-range SEI and collective dissipation. Both are governed by the symmetric collective lowering operator $\hat J_B^-=\sum_{j=1}^{N}\hat\sigma_j^-$, where $\hat\sigma_j^-=|g\rangle_j\langle e|_j$. The coherent interaction $V\hat J_B^+\hat J_B^-$ includes both spin exchange between emitters and collective energy shifts, with $V=-{g_B^2\Delta_b}/{\Delta_b^2+\kappa_b^2}$ and $\Gamma_B={g_B^2\kappa_b}/{\Delta_b^2+\kappa_b^2}$. Here $\Delta_b$ is the auxiliary-cavity detuning from the atomic transition, $\kappa_b$ is the auxiliary-cavity field decay rate, and $\Gamma_B$ characterizes the induced collective dissipation~\cite{ReiterSorensen2012,Ritsch2013}.  Then the resulting effective Hamiltonian reads ($\hbar=1$)
\begin{equation}
\begin{aligned}
\hat H_{\mathrm{eff}}={}&
\Delta_a\hat a^\dagger\hat a
+\delta\sum_{j=1}^{N}\hat\sigma_j^+\hat\sigma_j^-
+V\hat J_B^+\hat J_B^-\\
&+\hat a^\dagger\sum_{j=1}^{N}g_j\hat\sigma_j^-
+\mathrm{H.c.}
+i\eta(\hat a^\dagger-\hat a),
\end{aligned}
\label{eq:Heff}
\end{equation}
where $\Delta_a$ and $\delta$ are the primary-cavity and atomic detunings, respectively.

Including complete dissipations of system, the dynamics of density matrix is described by the master equation
\begin{equation}
\begin{aligned}
\dot{\hat\rho}={}&
-i[\hat H_{\mathrm{eff}},\hat\rho]
+2\kappa\mathcal D[\hat a]\hat\rho
+2\Gamma_B\mathcal D[\hat J_B^-]\hat\rho\\
&+\sum_{i,j=1}^{N}\gamma_{ij}
\bigl(
2\hat\sigma_i^-\hat\rho\hat\sigma_j^+
-\{\hat\sigma_j^+\hat\sigma_i^-,\hat\rho\}
\bigr),
\end{aligned}
\label{eq:master}
\end{equation}
where $\mathcal D[\hat O]\hat\rho=
\hat O\hat\rho\hat O^\dagger-
\{\hat O^\dagger\hat O,\hat\rho\}/2$ denotes the Lindblad type of dissipation and $\kappa$ is the primary-cavity field decay rate. To account for spatially correlated atomic decay, as in photonic-crystal waveguides (PCWs)~\cite{Douglas2015,Goban2015,Hood2016}, we adopt the Markovian decay matrix $\gamma_{ij}=\gamma_p\delta_{ij}+\gamma_{1\mathrm{D}}\cos[\phi_g(i-j)]$ for equally spaced emitters coupled to a bidirectional guided reservoir~\cite{AsenjoGarcia2017,Sheremet2023}. Here $\gamma_p$ and $\gamma_{1\mathrm{D}}$ denote the independent and guided-mode decay rates, respectively, and $\phi_g$ is the propagation phase between adjacent emitters, giving a single-emitter coherence decay rate $\gamma=\gamma_p+\gamma_{1\mathrm{D}}$. We neglect coherent waveguide-mediated exchange to isolate correlated dissipation, with $\phi$ and $\phi_g$ specifying the primary-cavity coupling profile and reservoir correlations, respectively.

We first consider two emitters ($N=2$), for which the excitation-selective mechanism is illustrated in Fig.~\ref{fig:model}(b). The collective atomic states are $|G\rangle=|gg\rangle$, $|E\rangle=|ee\rangle$, and $|S,A\rangle=(|eg\rangle\pm|ge\rangle)/\sqrt{2}$. Introducing $\hat S^-=(\hat\sigma_1^-+\hat\sigma_2^-)/\sqrt{2}$ and $\hat A^-=(\hat\sigma_1^--\hat\sigma_2^-)/\sqrt{2}$, the primary-cavity interaction takes the form $\hat a^\dagger(g_S\hat S^-+g_A\hat A^-)+\mathrm{H.c.}$, where $g_{S,A}=g(1\pm\cos\phi)/\sqrt{2}$. The correlated reservoir gives $\gamma_c=\gamma_{1\mathrm{D}}\cos\phi_g$, with collective coherence decay rates $\gamma_A=\gamma-\gamma_c$, $\gamma_S=\gamma+\gamma_c+2\Gamma_B$, and $\gamma_E=2(\gamma+\Gamma_B)$ for the states $|A\rangle$, $|S\rangle$, and $|E\rangle$, respectively. At $\phi=\pi$, $g_S=0$ and $g_A=\sqrt{2}g$, so the primary cavity couples only to the antisymmetric state within the single-excitation manifold. Since this state is decoupled from the auxiliary channel, its resonance frequency and intrinsic decay rate are unaffected by $V$ and $\Gamma_B$, leaving the corresponding linear cavity antiresonance unchanged~\cite{Sames2014,Filipp2011,Plankensteiner2017}. For $\phi\neq\pi$, the symmetric state also contributes to the linear response.

This protection follows directly from the collective selection rules $\hat J_B^-|A\rangle=0$, $\hat J_B^-|S\rangle=\sqrt{2}|G\rangle$, and $\hat J_B^-|E\rangle=\sqrt{2}|S\rangle$. The auxiliary-induced non-Hermitian term $\hat H_{B,\mathrm{nh}}=(V-i\Gamma_B)\hat J_B^+\hat J_B^-$ therefore leaves $|A\rangle$ unaffected but shifts $|S\rangle$ and $|E\rangle$ by $2(V-i\Gamma_B)$. At $\phi=\pi$, the primary cavity couples the $|G\rangle\leftrightarrow|A\rangle\leftrightarrow|E\rangle$ ladder, whereas the auxiliary channel connects $|E\rangle$ to $|S\rangle$ [Fig.~\ref{fig:model}(b)]. The auxiliary interaction thus modifies the complex two-excitation spectrum and two-photon pathways without altering the protected linear antiresonance. This separation is enforced by collective symmetry rather than a fine-tuned cancellation between coherent and dissipative contributions. Unless otherwise specified, we use $N=2$, $g=\gamma=\kappa$, $V=\kappa$, $\Gamma_B=0.05\kappa$, $\gamma_c=0.95\gamma$, $\eta=0.1\kappa$, and $\Delta_a=0$.

\begin{figure}[t]
\centering
\includegraphics[width=0.99\columnwidth]{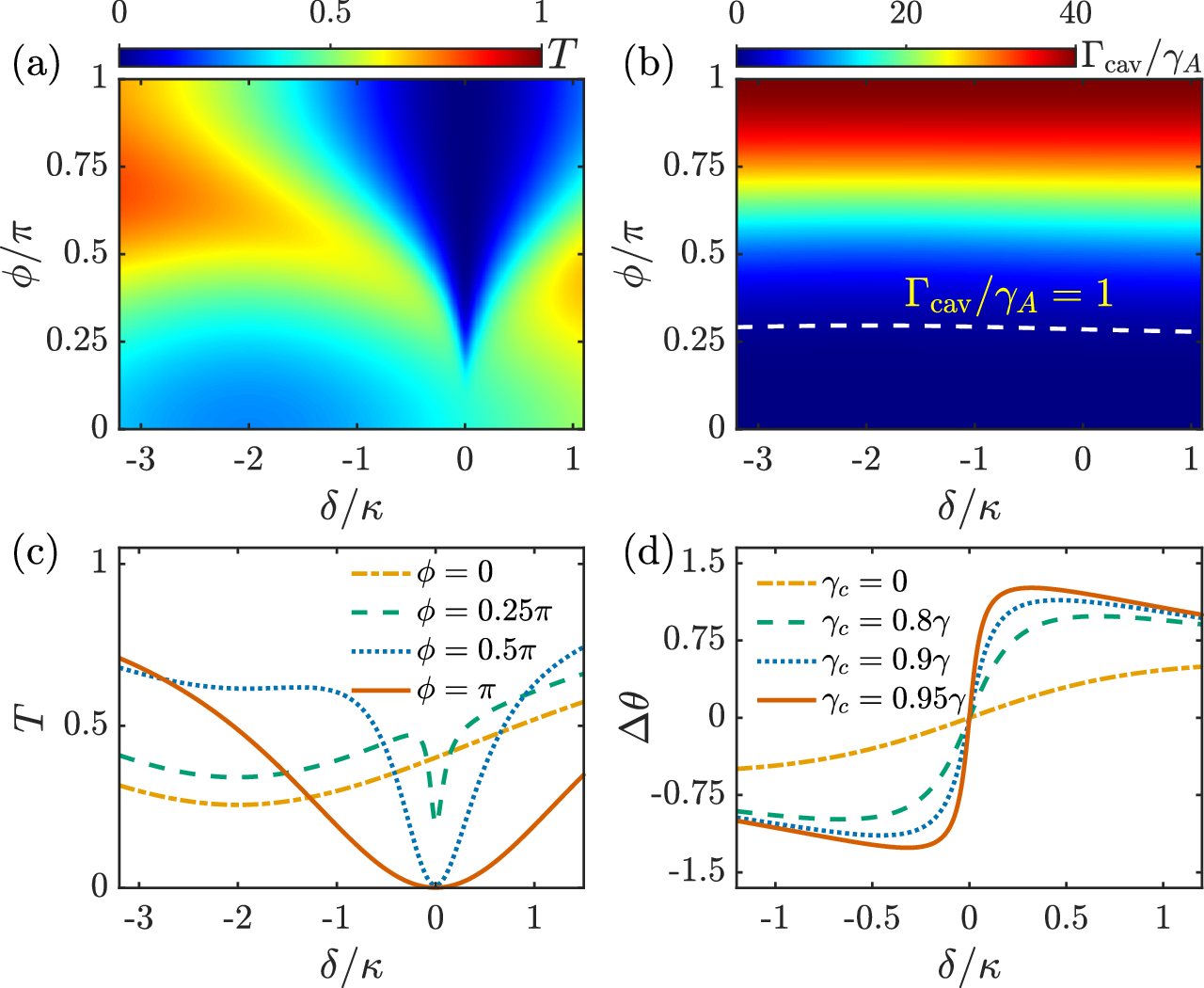}
\caption{\textbf{Phase-controlled subradiant antiresonance and optical response.} (a) Cavity transmission $T=|t|^2$ versus detuning $\delta$ and  phase $\phi$. (b) Cavity-induced decay $\Gamma_{\rm cav}$, versus $V$ and $\phi$. The dashed line marks $\Gamma_{\rm cav}=\gamma_A$. (c) Transmission spectra for different $\phi$. (d) Relative transmission phase $\Delta\theta=\theta(\delta)-\theta(0)$ at $\phi=\pi$ for different $\gamma_c$.}
\label{fig:antiresonance}
\end{figure}

{\em Symmetry-protected antiresonance and metrology.---}In the weak-drive limit, cavity transmission amplitude is $t(\delta,\phi)=\kappa/[\kappa+i\Delta_a+g_S^2/(\gamma_S+i\delta+2iV)+g_A^2/(\gamma_A+i\delta)]$. Figure~\ref{fig:antiresonance}(a) shows how phase $\phi$ controls the relative contributions of the symmetric and antisymmetric collective states to the cavity transmission. At $\phi=0$, only the symmetric state couples to  cavity, producing an antiresonance near $\delta=-2V$. As $\phi$ increases, the coupling to the antisymmetric state grows, giving rise to a second antiresonance near $\delta=0$ [Fig.~\ref{fig:antiresonance}(c)]. At $\phi=\pi$, the symmetric state decouples from cavity, and the transmission reduces to $t_\pi(\delta)=\kappa/[\kappa+i\Delta_a+2g^2/(\gamma_A+i\delta)]$. Its independence from $V$ and $\Gamma_B$ establishes a symmetry-protected antiresonance whose intrinsic resonance frequency and coherence decay rate remain unaffected by the auxiliary channel. The coupling phase thus tunes the cavity response without altering the intrinsic properties of the subradiant state~\cite{Sames2014,Plankensteiner2017}.

To quantify the distinction between intrinsic protection and tunable cavity coupling, we consider the self-energy of the antisymmetric state. Near $\delta=0$, the symmetric contribution provides a slowly varying background $K(\phi,V)=\kappa+i\Delta_a+g_S^2/(\gamma_S+2iV)$, provided the two collective resonances are sufficiently separated. The corresponding local self-energy is
\begin{equation}
\Sigma_A=\frac{g_A^2}{K}\equiv\Gamma_{\rm cav}-i\Delta_{\rm cav},
\label{eq:Sigma_readout}
\end{equation}
where $\Gamma_{\rm cav}=\operatorname{Re}\Sigma_A$ and $\Delta_{\rm cav}=-\operatorname{Im}\Sigma_A$ describe the cavity-induced coherence decay and frequency shift, respectively. Figure~\ref{fig:antiresonance}(b) shows the ratio $\Gamma_{\rm cav}/\gamma_A$, which characterizes the cavity-induced decay relative to intrinsic subradiant decay. For $\phi\ll1$, $g_A\approx g\phi^2/(2\sqrt{2})$, yielding $\Gamma_{\rm cav}\approx(g^2\phi^4/8)\operatorname{Re}(1/K_0)$, with $K_0=K(0,V)$. Thus, cavity-induced decay vanishes as $\phi^4$ near $\phi=0$, while $\gamma_A$ remains unchanged. This provides independent control over cavity-induced response while preserving the intrinsic subradiant decay rate.

The protected subradiant state also exhibits a pronounced transmission-phase response. Figure~\ref{fig:antiresonance}(d) shows that increasing $\gamma_c$ sharpens the phase dispersion near the antisymmetric antiresonance. At $\phi=\pi$ and $\Delta_a=0$, the transmission phase $\theta=\arg t$ has a central slope $\left.\partial_\delta\theta\right|_{\delta=0}=2g^2/[\gamma_A(\kappa\gamma_A+2g^2)]$. For $2g^2\gg\kappa\gamma_A$, the slope approaches $1/\gamma_A$, directly relating the phase response to the intrinsic decay rate of antisymmetric state. However, the steep phase dispersion occurs near suppressed transmission, so a large phase slope does not necessarily yield optimal measurement sensitivity. This motivates a Fisher-information analysis that accounts for both phase dispersion and transmitted photon flux.

\begin{figure}[t]
\centering
\includegraphics[width=0.99\columnwidth]{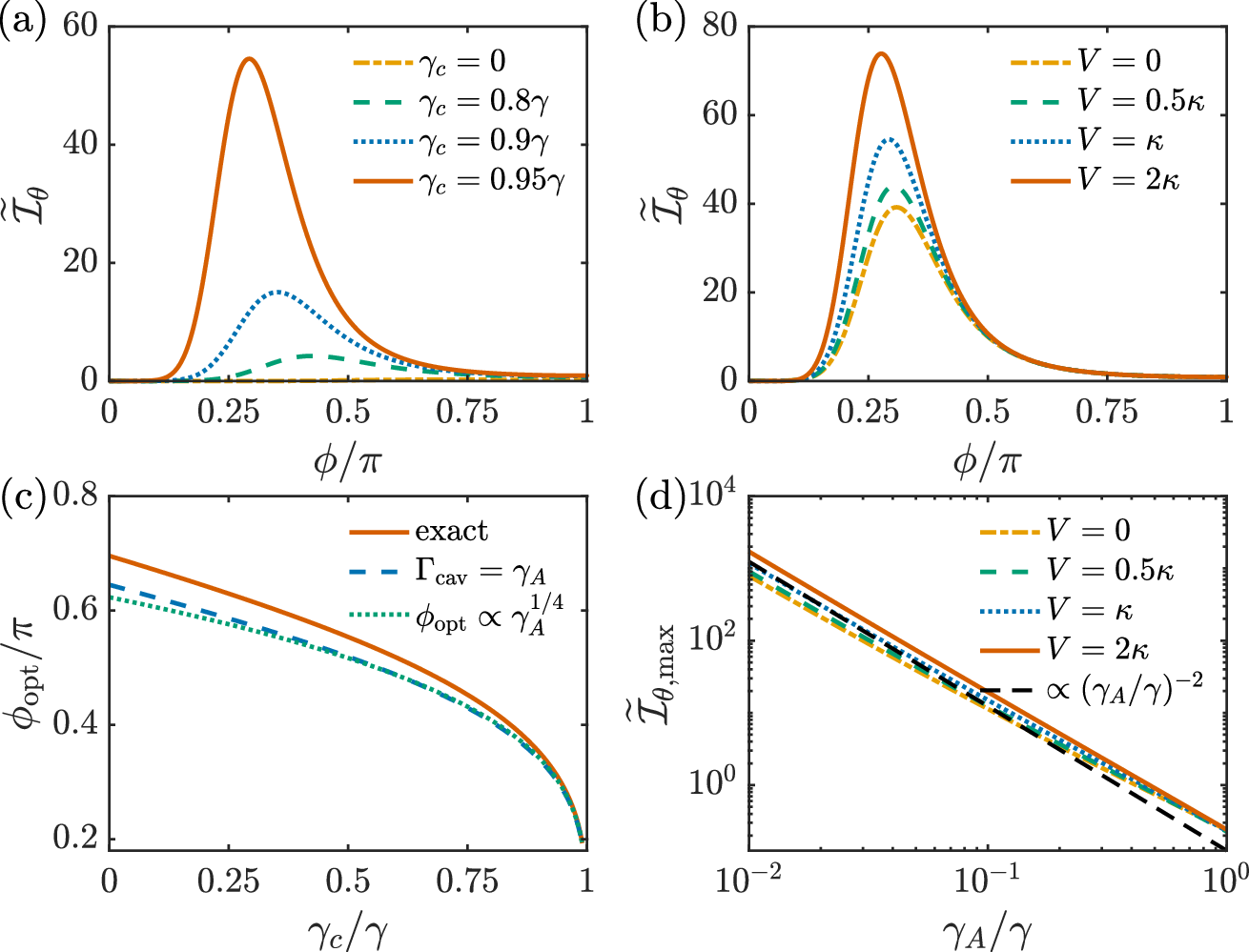}
\caption{\textbf{Phase Fisher information and optimal readout.} $\widetilde{\mathcal I}_{\theta}$ versus $\phi$ for different (a) $\gamma_c$ with $V/\kappa=1$ and (b) $V$ with $\gamma_c/\gamma=0.95$. (c) $\phi_{\rm opt}$ from full calculation (solid), readout-decay matching $\Gamma_{\rm cav}=\gamma_A$ (dashed), and asymptotic scaling $\phi_{\rm opt}\propto\gamma_A^{1/4}$ (dotted). (d) $\widetilde{\mathcal I}_{\theta}$ versus $\gamma_A$ for different $V$, with dashed line showing asymptotic inverse-square scaling.}
\label{fig:sensing}
\end{figure}

The steep phase dispersion of the protected antiresonance offers enhanced phase sensitivity, but the suppressed transmission limits the available signal. To quantify this trade-off, we consider a perturbation $\lambda$ that shifts the atomic detuning, $\delta(\lambda)=\delta_0+\chi_\lambda\lambda$, with $\chi_\lambda=\partial\delta/\partial\lambda$. For a coherent probe and ideal homodyne detection of the phase quadrature, the classical Fisher information per incident photon is $\mathcal I_{\lambda,\theta}^{(\mathrm{in})}=4T(\partial_\lambda\theta)^2$~\cite{Giovannetti2011,Demkowicz2015}. This expression accounts for the coherent phase response but excludes information from amplitude variations and output fluctuations. We introduce the dimensionless phase Fisher information
\begin{equation}
\widetilde{\mathcal I}_{\theta}
=\frac{\kappa^2}{\chi_\lambda^2}\mathcal I_{\lambda,\theta}^{(\mathrm{in})}
=4T\left|\frac{\partial\theta}{\partial(\delta/\kappa)}\right|^2.
\label{eq:Itheta_dimensionless}
\end{equation}
Unlike the phase slope alone, this quantity incorporates the loss of transmitted photons near the antiresonance. Figure~\ref{fig:sensing}(a) shows that the Fisher information is maximized at an intermediate coupling phase. Increasing $\gamma_c$ enhances this maximum while shifting the optimal phase toward weaker coupling to the antisymmetric state. The optimum thus reflects a balance between the intrinsic subradiant decay and cavity-induced optical readout.

To identify the origin of this optimum, we examine the transmission near the antisymmetric resonance. Treating the symmetric contribution as the slowly varying background $K$ defined above yields
\begin{equation}
t(\delta)\approx\frac{\kappa}{K}\frac{\gamma_A+i\delta}{\gamma_A+\Gamma_{\rm cav}+i(\delta-\Delta_{\rm cav})}.
\label{eq:t_local}
\end{equation}
For a negligible or compensated cavity-induced frequency shift, the on-resonance transmission and phase slope are $T_0\approx T_{\rm bg}[\gamma_A/(\gamma_A+\Gamma_{\rm cav})]^2$ and $\left.\partial_\delta\theta\right|_0\approx\Gamma_{\rm cav}/[\gamma_A(\gamma_A+\Gamma_{\rm cav})]$, respectively, where $T_{\rm bg}=|\kappa/K|^2$. The resulting phase Fisher information is
\begin{equation}
\frac{\mathcal I_{\lambda,\theta}^{(\mathrm{in})}}{\chi_\lambda^2}
\approx4T_{\rm bg}\frac{\Gamma_{\rm cav}^2}{(\gamma_A+\Gamma_{\rm cav})^4}.
\label{eq:FI_local}
\end{equation}
For a fixed optical background, maximizing this expression with respect to $\Gamma_{\rm cav}$ gives $\Gamma_{\rm cav}=\gamma_A$. The optimal phase information therefore occurs when the cavity-induced coherence decay matches the intrinsic subradiant decay, rather than when the optical coupling or phase slope is maximized. This readout-decay matching condition provides a local criterion for converting protected subradiant coherence into measurable phase information.

The auxiliary interaction further controls the readout-decay balance. Figure~\ref{fig:sensing}(b) shows that varying $V$ shifts the information maximum through the symmetric-state contribution to $K$, without affecting the protected antisymmetric state. In the strongly subradiant regime, $\phi_{\rm opt}\to0$, where $\Gamma_{\rm cav}\approx(g^2\phi^4/8)\operatorname{Re}(1/K_0)$. The matching condition $\Gamma_{\rm cav}=\gamma_A$ then yields optimal phase
\begin{equation}
\phi_{\rm opt}\approx
\left[\frac{8\gamma_A}{g^2\operatorname{Re}(1/K_0)}\right]^{1/4}
\propto\gamma_A^{1/4}.
\label{eq:phi_opt_scaling}
\end{equation}
This fourth-root scaling follows from $g_A\propto\phi^2$ near $\phi=0$. Figure~\ref{fig:sensing}(c) compares this prediction with the full transmission calculation, showing agreement within the validity of the local approximation.

At the matching point, the transmission remains finite, $T_0\approx T_{\rm bg}/4$, while $|\partial_\delta\theta|_0\approx1/(2\gamma_A)$, giving $\mathcal I_{\lambda,\theta,\max}^{(\mathrm{in})}\approx\chi_\lambda^2T_{\rm bg}/(4\gamma_A^2)$. For a finite, nonzero background transmission and a valid local approximation as $\gamma_A\to0$, the optimized dimensionless phase information scales as $\widetilde{\mathcal I}_{\theta,\max}\propto(\gamma_A/\gamma)^{-2}$ for $\gamma_A/\gamma\ll1$, with other frequency scales fixed. Figure~\ref{fig:sensing}(d) illustrates this scaling, with $V$ modifying the prefactor through the optical background. Thus, phase-controlled readout converts long-lived subradiant coherence into enhanced phase information while maintaining finite transmission.

\begin{figure}[t]
\centering
\includegraphics[width=0.99\columnwidth]{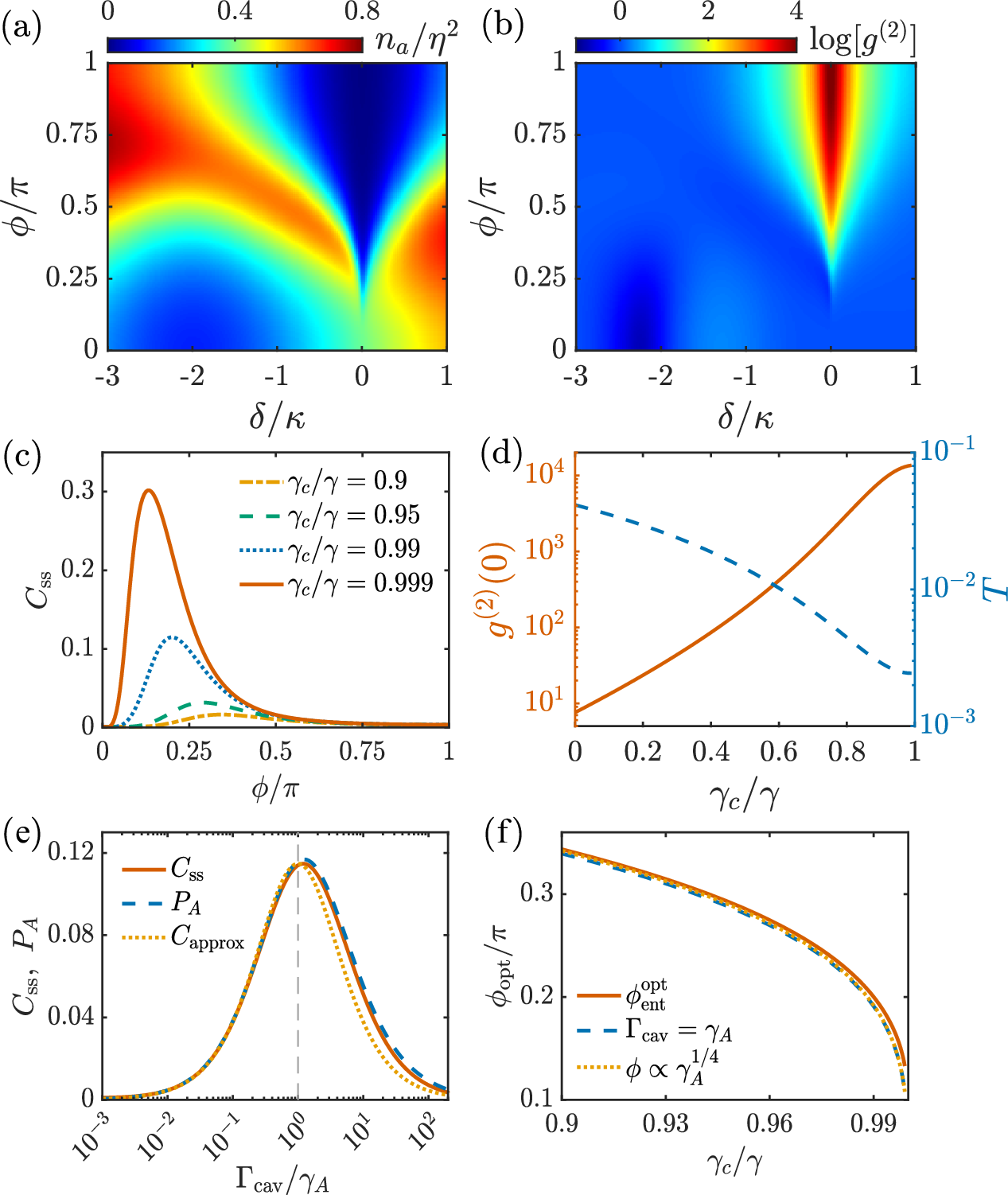}
\caption{\textbf{Photon correlations and atomic entanglement.} (a) $n_a/\eta^2$ and (b) $\log g^{(2)}(0)$ versus $\delta$ and $\phi$. (c) Concurrence $C_{\rm ss}$ versus $\phi$ for different $\gamma_c/\gamma$ at $\delta=0$. (d) $g^{(2)}(0)$ (solid) and $T$ (dashed) versus $\gamma_c/\gamma$ at $(\delta,\phi)=(0,\pi)$. (e) $C_{\rm ss}$ (solid) and  $P_A$ (dashed) versus $\Gamma_{\rm cav}$ at $\gamma_c/\gamma=0.99$. (f) $\phi_{\rm ent}^{\rm opt}$ from the master equation (solid), readout-decay matching (dashed), and fourth-root scaling (dotted).}
\label{fig:quantum_correlations}
\end{figure}

{\em Photon correlations and atomic entanglement.---}We next examine how symmetry protection shapes photon correlations and steady-state atomic entanglement beyond the linear response. From the steady-state master equation, we calculate the intracavity population $n_a=\langle\hat a^\dagger\hat a\rangle$ and second-order correlation $g^{(2)}(0)=\langle\hat a^{\dagger 2}\hat a^2\rangle_{\rm ss}/n_a^2$. Figures~\ref{fig:quantum_correlations}(a) and \ref{fig:quantum_correlations}(b) show pronounced photon bunching near the antisymmetric antiresonance, accompanied by a strongly suppressed cavity population. The large $g^{(2)}(0)$ reflects the relative survival of two-photon excitation pathways but does not necessarily imply an enhanced photon-pair flux.

This contrast originates from the distinct action of the auxiliary collective channel in different excitation manifolds. At $\phi=\pi$, the single-excitation pathway $|0,G\rangle\leftrightarrow|1,G\rangle\leftrightarrow|0,A\rangle$ involves the protected state $|A\rangle$, satisfying $\hat J_B^-|A\rangle=0$. The auxiliary channel therefore leaves the linear antiresonance unchanged. In contrast, the two-excitation pathway $|2,G\rangle\leftrightarrow|1,A\rangle\leftrightarrow|0,E\rangle$ involves $|E\rangle=|ee\rangle$, for which $\hat J_B^-|E\rangle=\sqrt{2}|S\rangle$. Its diagonal complex energy in the no-jump effective Hamiltonian is $\varepsilon_E=2\delta+2V-2i(\gamma+\Gamma_B)$. The auxiliary channel thus modifies two-photon excitation pathways without shifting the protected linear resonance, allowing strong bunching near suppressed transmission, in contrast to conventional photon blockade~\cite{Birnbaum2005}. Figure~\ref{fig:quantum_correlations}(d) further shows that increasing correlated decay enhances $g^{(2)}(0)$ while reducing transmission.

Beyond photon correlations, the protected antisymmetric state also supports steady-state atomic entanglement. We quantify the entanglement of reduced atomic state $\hat\rho_{\rm at}=\operatorname{Tr}_{\rm cav}\hat\rho_{\rm ss}$ using Wootters concurrence $C_{\rm ss}$~\cite{Wootters1998}. Figure~\ref{fig:quantum_correlations}(c) shows that increasing correlated decay enhances $C_{\rm ss}$, which peaks at an intermediate coupling phase rather than at $\phi=\pi$. As shown in Fig.~\ref{fig:quantum_correlations}(e), the concurrence closely follows the antisymmetric-state population $P_A=\langle A|\hat\rho_{\rm at}|A\rangle$, identifying the antisymmetric Bell state as the dominant entangled component. Indeed, for states restricted to the $\{|G\rangle,|A\rangle\}$ subspace, $C=P_A$, even in the presence of coherence between these states.

The finite-phase entanglement maximum can be understood from the weak-excitation population of the antisymmetric state, $P_A\approx\eta^2g_A^2/|K(\gamma_A+i\delta)+g_A^2|^2$. Using $g_A^2/K=\Gamma_{\rm cav}-i\Delta_{\rm cav}$, we obtain $P_A\approx(\eta^2/\operatorname{Re}K)\Gamma_{\rm cav}/[(\gamma_A+\Gamma_{\rm cav})^2+(\delta-\Delta_{\rm cav})^2]$ for $\operatorname{Re}K>0$. For negligible effective detuning and a slowly varying background, $C_{\rm ss}\approx P_A\propto\Gamma_{\rm cav}/(\gamma_A+\Gamma_{\rm cav})^2$, yielding an optimum at $\Gamma_{\rm cav}=\gamma_A$. This approximate matching condition accounts for the finite-phase maximum in Fig.~\ref{fig:quantum_correlations}(e).

Figure~\ref{fig:quantum_correlations}(f) compares the optimal entanglement phase from the full master equation with the weak-excitation prediction. In the strongly subradiant regime, $\phi_{\rm ent}^{\rm opt}\to0$, where $g_A^2\approx g^2\phi^4/8$ and $K\approx K_0$. At $\delta=0$, maximizing the antisymmetric-state population yields $g_{A,\rm opt}^2\approx\gamma_A|K_0|$ and hence $\phi_{\rm ent}^{\rm opt}\approx(8\gamma_A|K_0|/g^2)^{1/4}\propto\gamma_A^{1/4}$. This reproduces the fourth-root scaling found for optimal phase sensing, although the two optimal phases generally differ because they obey distinct optimization conditions. Saturation and variations in the optical background can lead to deviations from the weak-excitation prediction. These results show that symmetry protection preserves the linear antiresonance while enabling tunable photon correlations, atomic entanglement, and phase-sensitive optical readout.

\begin{figure}[t]
\centering
\includegraphics[width=0.99\columnwidth]{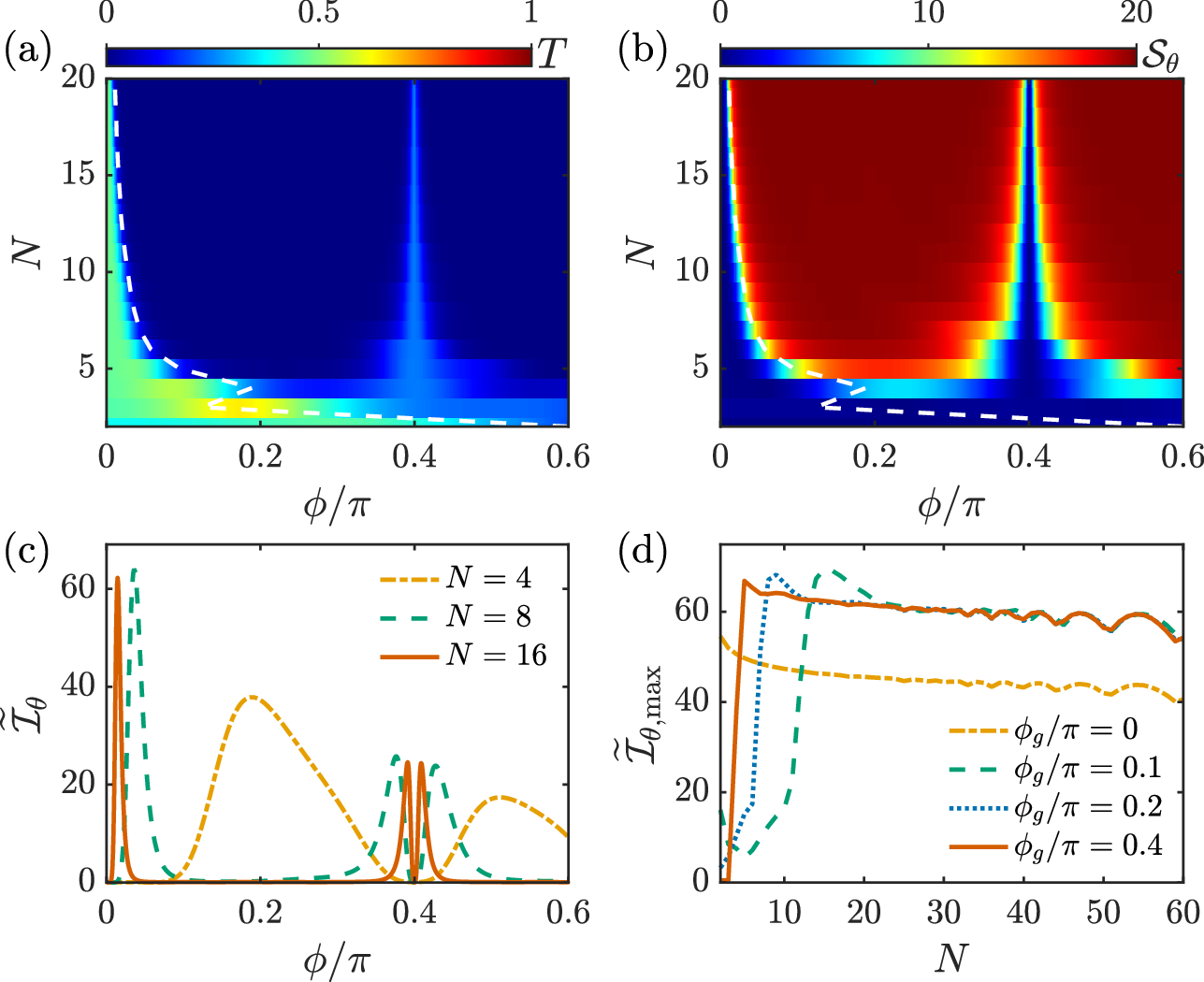}
\caption{\textbf{Phase-sensitive optical response of correlated emitter arrays.} (a) Transmission $T$ and (b) phase susceptibility $\mathcal S_\theta$ versus coupling phase $\phi$ and emitter number $N$. White dashed curves indicate the optimal phase $\phi_{\rm opt}(N)$ maximizing $\widetilde{\mathcal I}_\theta=4T\mathcal S_\theta^2$. (c) Phase Fisher information $\widetilde{\mathcal I}_\theta$ versus $\phi$ for $N=4$, $8$, and $16$. (d) Optimized phase Fisher information $\widetilde{\mathcal I}_{\theta,\max}$ versus $N$ for the indicated reservoir phases $\phi_g$. Panels (a)--(c) use $\phi_g=0.4\pi$. Other parameters are $\gamma_{1\mathrm{D}}=0.95\gamma$ and $\gamma_p=0.05\gamma$.}
\label{fig:array}
\end{figure}

{\em Collective subradiant response in emitter arrays.---}Extending the linear-response analysis to correlated emitter arrays, we find that spatially structured cavity coupling provides tunable optical access to protected collective modes. The two-emitter results show that subradiance-enhanced phase sensing requires both long-lived collective coherence and finite optical coupling. We now examine how this interplay extends to emitter arrays, where multiple collective modes can contribute to the cavity response. For $N$ equally spaced emitters, the main-cavity coupling acquires the spatial profile $g_j=g\cos[(j-1)\phi]$, while the auxiliary channel remains uniform, $\hat J_B^-=\sum_{j=1}^N\hat\sigma_j^-$. The independent phases $\phi_g$ and $\phi$ then control the collective decay structure and the optical accessibility of its eigenmodes, respectively.

In the weak-excitation regime, the atomic coherences $\mathbf s=(\langle\hat\sigma_1^-\rangle,\ldots,\langle\hat\sigma_N^-\rangle)^T$ satisfy $\mathbf M(\delta)\mathbf s=-i\mathbf g\alpha$, where $\alpha=\langle\hat a\rangle$. Eliminating the atomic coherences yields
\begin{equation}
\begin{aligned}
\mathbf M(\delta)&=\boldsymbol{\Gamma}+(\Gamma_B+iV)\mathbf u\mathbf u^T+i\delta\mathbf I,\\
t(\delta,\phi)&=\frac{\kappa}{\kappa+i\Delta_a+\mathbf g^T\mathbf M^{-1}(\delta)\mathbf g},
\end{aligned}
\label{eq:array_response}
\end{equation}
where $\mathbf u=(1,\ldots,1)^T$ and $\mathbf g=(g_1,\ldots,g_N)^T$. The correlated decay matrix can be written as $\boldsymbol{\Gamma}=\gamma_p\mathbf I+\gamma_{1D}(\mathbf c\mathbf c^T+\mathbf s_g\mathbf s_g^T)$, with $c_j=\cos[(j-1)\phi_g]$ and $s_{g,j}=\sin[(j-1)\phi_g]$. Since the guided-mode contribution has rank at most two, at least $N-2$ collective modes retain only the residual decay $\gamma_p$. Imposing orthogonality to the uniform auxiliary channel further defines the protected subspace $\mathcal D=\ker[\mathbf c,\mathbf s_g,\mathbf u]^T$, whose dimension is $N-\operatorname{rank}[\mathbf c,\mathbf s_g,\mathbf u]$, generically $N-3$. For any $\mathbf v\in\mathcal D$, $\mathbf M(\delta)\mathbf v=(\gamma_p+i\delta)\mathbf v$, showing that these modes are unaffected by both the auxiliary-induced coherent interaction and collective decay.

The existence of protected collective modes does not, however, guarantee their optical accessibility. Introducing the orthogonal projector $\mathbf P_D$ onto $\mathcal D$, the protected contribution to the cavity self-energy is
\begin{equation}
\chi_D(\delta,\phi)=\frac{\mathbf g^T\mathbf P_D\mathbf g}{\gamma_p+i\delta}.
\label{eq:array_dark_response}
\end{equation}
The optical weight $W_D(\phi)=\mathbf g^T\mathbf P_D\mathbf g=\|\mathbf P_D\mathbf g\|^2$ quantifies the coupling of the cavity field to the protected subspace. In particular, $W_D=0$ when $\mathbf g$ lies in the span of $\mathbf c$, $\mathbf s_g$, and $\mathbf u$, rendering these long-lived modes optically dark. Varying $\phi$ therefore controls their optical accessibility without changing their intrinsic linewidths, while the remaining collective modes contribute an additional optical background. This extends the separation between subradiant protection and optical readout from a single antisymmetric state to a collective subspace.

Figures~\ref{fig:array}(a) and \ref{fig:array}(b) show the resonant transmission $T=|t|^2$ and phase susceptibility $\mathcal S_\theta=|\partial\theta/\partial(\delta/\kappa)|_{\delta=0}$ as functions of $\phi$ and $N$, where $\theta=\arg t$. Their phase dependence reflects the redistribution of optical coupling among collective modes with different decay rates. Defining $D_c=\kappa+i\Delta_a+\mathbf g^T\mathbf M^{-1}\mathbf g$ and using $\partial_\delta\mathbf M^{-1}=-i\mathbf M^{-2}$, we obtain
\begin{equation}
\mathcal S_\theta=\kappa\left|\operatorname{Re}\left[\frac{\mathbf g^T M^{-2}\mathbf g}{D_c}\right]\right|_{\delta=0}.
\label{eq:array_phase_slope}
\end{equation}
As in the two-emitter system, strong phase dispersion can be accompanied by suppressed transmission. We therefore characterize the phase sensitivity by the phase Fisher information per incident photon, $\widetilde{\mathcal I}_\theta=4T\mathcal S_\theta^2$, which accounts for the transmission penalty. The dashed curves in Figs.~\ref{fig:array}(a) and \ref{fig:array}(b) mark the phases that maximize $\widetilde{\mathcal I}_\theta$ within the selected interval.

Figure~\ref{fig:array}(c) shows that $\widetilde{\mathcal I}_\theta(\phi)$ develops multiple local maxima as $N$ increases, reflecting the phase-dependent coupling to different collective response channels. Unlike the two-emitter system, the array generally involves several collective modes and a nontrivial optical background. Consequently, the single-mode matching condition $\Gamma_{\rm cav}=\gamma_A$ does not universally determine the optimal phase. Figure~\ref{fig:array}(d) further shows that substantial optimized phase information persists for arrays of up to $N=60$, with a pronounced dependence on the reservoir phase $\phi_g$. The optimization is restricted to $0<\phi/\pi\leq0.6$; the resulting size dependence does not imply universal collective enhancement. These results extend the separation between subradiant protection and optical accessibility from a single antisymmetric excitation to a collective subspace. While reservoir correlations establish long-lived collective modes, spatially selective cavity coupling determines their contribution to the measured phase response. 

{\em Conclusion.---}We have demonstrated that collective symmetry enables independent control of nonlinear excitation pathways and optical readout while preserving a subradiant cavity antiresonance. The auxiliary collective channel decouples from the antisymmetric single-excitation state but remains active in the two-excitation manifold, allowing photon correlations to be tuned without shifting or broadening the protected linear resonance. Independently, the cavity coupling phase controls optical access to the subradiant state without altering its intrinsic decay rate. This separation enables pronounced photon bunching and tunable atomic entanglement, while the competition between optical readout and subradiant decay yields an optimal phase Fisher information per incident photon. This phase-selective readout mechanism further extends to protected collective modes in correlated emitter arrays.

The protection stems from the symmetry of the auxiliary collective coupling rather than a cancellation of coherent and dissipative effects. Frequency disorder and nonuniform auxiliary couplings can mix bright and dark states, while additional radiative losses broaden the subradiant resonance. Coherent photon-mediated exchange can further reshape the collective spectrum~\cite{AsenjoGarcia2017,Sheremet2023}. Recent studies of collective photon correlations and multiexcitation subradiance highlight the importance of nonlinear excitation pathways~\cite{Singh2025,Williamson2020,RubiesBigorda2023,Tecer2024}. Extending excitation-selective protection to higher-excitation manifolds in larger arrays remains an important direction for controlling multiphoton correlations.

{\em Acknowledgments}.---This work was supported by the National Natural Science Foundation of China (Grant No. 12674338 and Grant No. 12374365), Quantum Science and Technology-National Science and Technology Major Project (Grant No. 2025ZD0300400) and Guangdong Provincial Quantum Science Strategic Initiative (Grant Nos. GDZX2505001 and GDZX2604004).

%
%

\end{document}